\documentstyle[a4,12pt,epsf]{article}

\begin{document}

 \parindent 0pt \parskip \medskipamount

\newcommand{\Lslash}[1]{ \parbox[b]{1em}{$#1$} \hspace{-0.8em}
                         \parbox[b]{0.8em}{ \raisebox{0.2ex}{$/$} }    }
\newcommand{\Sslash}[1]{ \parbox[b]{0.6em}{$#1$} \hspace{-0.55em}
                         \parbox[b]{0.55em}{ \raisebox{-0.2ex}{$/$} }    }

\newcommand{\beq}{\begin{equation}}
\newcommand{\eeq}{\end{equation}}
\newcommand{\beqa}{\begin{eqnarray}}
\newcommand{\eeqa}{\end{eqnarray}}
\newcommand\myappendixsection{\setcounter{equation}{0} \section}
\newcommand\mysection{\setcounter{equation}{0}\section}
\renewcommand{\theequation}{\thesection.\arabic{equation}}
\newcounter{hran} \renewcommand{\thehran}{\thesection.\arabic{hran}}
\def\bmini{\setcounter{hran}{\value{equation}}
  \refstepcounter{hran}\setcounter{equation}{0}
  \renewcommand{\theequation}{\thehran\alph{equation}}\begin{eqnarray}}
\def\bminiG#1{\setcounter{hran}{\value{equation}}
\refstepcounter{hran}\setcounter{equation}{-1}
\renewcommand{\theequation}{\thehran\alph{equation}}
\refstepcounter{equation}\label{#1}\begin{eqnarray}}
\def\emini{\end{eqnarray}\relax\setcounter{equation}{\value{hran}}%
\renewcommand{\theequation}{\thesection.\arabic{equation}}}
\newcommand{\half}{\frac{1}{2}}
\newcommand{\gsim}{\buildrel > \over {_\sim}}
\newcommand{\lsim}{\buildrel < \over {_\sim}}
\newcommand{\ie}{{\it ie}}
\newcommand{\eg}{{\it eg}}
\newcommand{\cf}{{\it cf}}
\newcommand{\etal}{{\it et al.}}
\newcommand{\gev}{{\rm GeV}}
\newcommand{\tr}{{\rm Tr}}
\newcommand{\M}{{\cal M}}
\newcommand{\Pel}{{\cal P}_{el}}
\newcommand{\ieps}{i\epsilon\,}
\newcommand{\jpsi}{J/\psi}
\newcommand{\order}[1]{${\cal O}(#1)$}
\newcommand{\morder}[1]{{\cal O}(#1)}
\newcommand{\eq}[1]{Eq.\ (\ref{#1})}
\newcommand{\ktr}{k_\perp}
\newcommand{\ptr}{p_\perp}
\newcommand{\pvec}{\vec p}
\newcommand{\Pvec}{\vec P}
\newcommand{\kvec}{\vec k}
\newcommand{\qvec}{\vec q}
\newcommand{\xvec}{\vec x}
\newcommand{\qo}{q^0}
\newcommand{\kz}{k^z}
\newcommand{\epsla}{\Sslash{\varepsilon}}
\newcommand{\esla}{\Sslash{e}}
\newcommand{\psla}{\Sslash{p}}
\newcommand{\ksla}{\Sslash{k}}

\newcommand{\as}{\alpha_s}
\newcommand{\lqcd}{\Lambda_{QCD}}
\newcommand{\ket}[1]{\vert{#1}\rangle}
\newcommand{\bra}[1]{\langle{#1}\vert}
\newcommand{\ave}[1]{\langle{#1}\rangle}
\newcommand{\qpair}{q\bar q}
\newcommand{\cpair}{c\bar c}
\newcommand{\bpair}{b\bar b}
\newcommand{\dis}[1]{\displaystyle{#1}}

\newfont{\secf}{cmb10 scaled \magstep2}

\newcommand{\PL}[3]{Phys.\ Lett.\ {{\bf#1}} ({#2}) {#3}}
\newcommand{\NP}[3]{Nucl.\ Phys.\ {{\bf#1}} ({#2}) {#3}}
\newcommand{\PR}[3]{Phys.\ Rev.\ {{\bf#1}} ({#2}) {#3}}
\newcommand{\PRL}[3]{Phys.\ Rev.\ Lett.\ {{\bf#1}} ({#2}) {#3}}
\newcommand{\ZP}[3]{Z. Phys.\ {{\bf#1}} ({#2}) {#3}}
\newcommand{\PRe}[3]{Phys.\ Rep.\ {{\bf#1}} ({#2}) {#3}}

\begin{titlepage}
\begin{flushright}
        NORDITA--97/37 P\\
        revised\\
        \today
\end{flushright}

\vskip 2.5cm

\centerline{\Large \bf Rescattering Effects in Quarkonium
Production\footnote{Work supported in part by the EU/TMR contract ERB
FMRX-CT96-0008.}}

\vskip 1.5cm

\centerline{\bf Paul Hoyer and St\'ephane Peign\'e}
\centerline{\sl Nordita}
\centerline{\sl Blegdamsvej 17, DK--2100 Copenhagen \O, Denmark}

\vskip 2cm

\begin{abstract}
We study  $\eta_c$ and $\jpsi$  hadroproduction induced by multiple
scattering off fixed centres in the target. We determine the minimum
number of hard scatterings required
and show that additional soft scatterings may be
factorized, at the level of the  production amplitude for the $\eta_c$ and of
the cross section for the $\jpsi$.
The $\jpsi$ provides an interesting example of soft rescattering effects
occurring inside a hard vertex.
We also explain the qualitative
difference between the transverse  momentum broadening of the $\jpsi$ and of
the $\Upsilon$ observed in collisions on nuclei. We point out that
rescattering from spectators produced by beam and target parton evolution may
have important effects in $\jpsi$ production.
\end{abstract}

\end{titlepage}

\newpage
\renewcommand{\thefootnote}{\arabic{footnote}}
\setcounter{footnote}{0}
\setcounter{page}{1}

\mysection{Introduction}

Quarkonium production is a sensitive measure of soft rescattering effects in
hard collisions. Whereas the creation of a heavy $Q\bar Q$ pair is a
process of scale $m_Q$, the binding energy of the heavy quarks in
non-relativistic quarkonium is only on the order of $\as^2\,m_Q$. Thus
for charmonium
the binding energy is of order $2m_D-m_{\jpsi} \simeq 600$ MeV,
which is only moderately larger than $\Lambda_{QCD}$.

There is considerable debate concerning the correct theoretical description of
quarkonium production. Whereas the `color singlet mechanism'
(CSM) \cite{csm1} appears to give good agreement with $\jpsi$
photoproduction data \cite{csm2}, it
fails by more than one order of magnitude in hadroproduction
\cite{schulerrev,sansoni,mangano,vhbt}. This has motivated
proposals of production mechanisms where radiation at the binding energy
scale plays an essential role, such as `color evaporation'
\cite{gavai,schuler,amundson,schulvogt} and the `color octet mechanism'
\cite{cgmp,com,annrev}. All models seem to have problems with some aspects of
the data, however.

The observed nuclear target $A$-dependence \cite{hoyrev} of the quarkonium
cross section gives further insight into the production process. Compared to
lepton pair production at the same hardness, quarkonium production shows
much bigger nuclear effects. The cross sections
(averaged over $x_F \gsim 0$)
can be parametrized as
\beq
\sigma_A \simeq \sigma_N \, A^{\alpha}
\,,\ \ \ \mbox{with}\
\left\{ \begin{array}{cc}
\alpha \simeq 1.00&\mu^+\mu^- \\
\alpha \simeq 0.92\pm 0.008&\jpsi, \psi' \\
\alpha \simeq 0.962\pm 0.014&\Upsilon(1S) \\
\end{array} \right.  \label{sigadep}
\eeq
These values are taken from Refs. \cite{E772dy,E772jpsi,E772upsilon},
respectively.

Furthermore, there is evidence that the
$\ptr$-broadening induced by the nucleus depends on the quark mass
\cite{E772upsilon,NA3},
\beq
\ave{\ptr^2(A)}-\ave{\ptr^2(^2H)} =
\left\{ \begin{array}{ccc}
0.113\pm 0.016 &\gev^2&\mu^+\mu^- \\
0.34\pm 0.08 &\gev^2&\jpsi, \psi' \\
0.667\pm 0.133 &\gev^2&\Upsilon (1S) \\
\end{array} \right.
\label{ptadep}
\eeq
where $A=184$ (W) for the E772 data on $\mu^+\mu^-$ and $\Upsilon$ production,
whereas $A=195$ (Pt) for the NA3 data on $\jpsi$ production.

The sizeable nuclear effects in quarkonium production shown by Eqs.
(\ref{sigadep}) and (\ref{ptadep}) raise two questions, which we shall
consider in this paper.

\begin{itemize}
\item[(i)]
Can $^3$S$_1$ quarkonia be produced through a hard
$gg \to Q\bar Q$ subprocess, with an additional {\em soft} rescattering gluon
coupled to the $Q\bar Q$ pair? If not, how does soft rescattering factorize
from the hard production amplitude?
\item[(ii)] A color octet
heavy $Q\bar Q$ pair of high momentum should behave like a
pointlike gluon in soft rescattering processes. How can the
$\ptr$-broadening depend on the quark mass, as indicated by \eq{ptadep}?
\end{itemize}

These questions concern the interplay of soft rescattering and hard
production processes, which have so far not been thoroughly studied in QCD
\cite{sterman}.
Soft rescattering by itself
has been studied in the problems  of high energy parton propagation
and energy loss in a dense or hot medium \cite{energyloss}.

Hard production amplitudes have been calculated at
leading twist, involving only a single scatterer in the target. A
factorization between hard and soft processes  \cite{CSS}
is usually assumed to hold also in quarkonium production
-- which is
indeed necessary for any firm perturbative predictions. The puzzles of
quarkonium production and the experimental evidence for rescattering effects
that depend on the hard scale motivate a more detailed investigation.

We shall study the rescattering and production of quarkonium in the limit of
(asymptotically) high energy. Since the hardness of the process (as measured
by the quark mass) is kept fixed, this implies that the
momentum transfer to
the target is a (vanishingly) small fraction of the projectile energy. Such a
limit seems natural for much of the data, which is at high energy and appears
to obey Feynman scaling.

The high energy limit considerably simplifies
elementary cross sections. For ex\-am\-ple, it is easy to verify that the
$e\mu \to e\mu$ cross section reduces to the Rutherford one when the
electron momentum tends to infinity while the transverse momentum transfer
and the
muon momentum are kept fixed. The fact that the scattering cross section in
this limit is independent of the target momentum and mass appears to be quite
general, and holds at least to lowest order for elementary targets.

We shall take advantage of this simplification by modelling the target partons
by very heavy quarks. This suppresses energy transfer and selects Coulomb
exchange, thus considerably simplifying the calculation. In the Appendix we
show explicitly (using the $\gamma g\to \qpair$ process) how the number
of heavy quarks in our kinematic limit is related to the gluon distribution
$G(x\to 0)$. A similar approach has earlier been used for describing deep
inelastic lepton scattering in the target rest frame and the creation of
rapidity gaps \cite{BHM}.

Rescattering effects in quarkonium and Drell-Yan production have
been con\-si\-dered previously, see, \eg, Ref. \cite{rescatteffects}.
Our assumption of a simple target structure allows a systematic and precise
investigation of the multiple scattering amplitudes in QCD. This
reveals interesting aspects of factorization between hard and soft processes,
color dynamics and $p_{\perp}$-broadening.

We consider processes of the type shown in Fig. \ref{genfig}. An incoming
gluon of asymptotically high energy mutiply
scatters in the target producing a heavy quarkonium. We use charm
to represent a typical heavy quark
and consider both $^1$S$_0$ $(\eta_c)$ and $^3$S$_1$ $(\jpsi)$ production. We
assume that the bound state is produced in a color singlet state, \ie,
the $\cpair$ pair couples directly
to the charmonium through its wave function at the origin.
Thus we do not consider higher order terms in an expansion in the relative
velocity of the charm quarks, which are relevant for the color octet mechanism
\cite{cgmp,com,annrev}.
Since we do not
consider gluon radiation, at least two target scatterings are required by
charge conjugation invariance
to produce a $\jpsi$, while one is sufficient for the $\eta_c$. We are
interested in understanding the systematics between hard and soft
Coulomb gluon exchanges.

\begin{figure}[htb]
\centerline{\ \ \ \ \ \ \ \ \ \ \ \ \ \ \ \ \ \
\vbox{\epsfxsize=14.5truecm\epsfbox{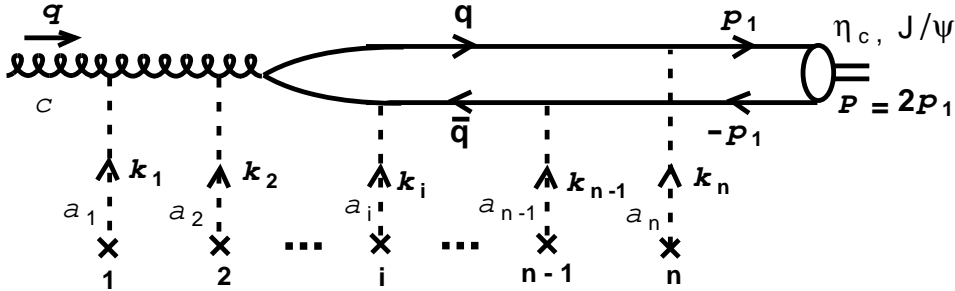}}}
\caption[*]{General amplitude for $\eta_c$ or $\jpsi$ production
induced by $n$ scatterings off static centres $1,\ldots\,n$ located at
$\xvec_1,\ldots\,\xvec_n$. The static centres are ordered and labelled
according to their increasing longitudinal position, and $\kvec_i$
is by definition the momentum transfered to the $\qpair$ pair by centre
\#  $i$. The color indices of the incident gluon and of the exchanged gluons
are denoted by $c$ and $a_i$, respectively.}
\label{genfig}
\end{figure}
A QED illustration will define more precisely what we mean by `hard' and
`soft' momentum transfers in this work. The
Rutherford scattering $e+A \to e+A$ of a {\em charged} particle off a heavy
nucleus is described by the Feynman diagram of Fig. \ref{qedfig}a. Due to the
Coulomb photon propagator the amplitude is proportional to $1/\ktr^2$, where
$\ktr$ is the transverse momentum transfer. The cross section is then given
by an
infrared divergent integral, $\sigma \propto \int d^2 \vec \ktr/\ktr^4$. For
scattering on neutral atoms, the infrared cutoff is given by the inverse
atomic radius $R$,
\beq
\mbox{$\ktr$ is soft} \Longleftrightarrow \ktr \sim \frac{1}{R}
\label{ksoft}
\eeq
On the other hand, in the pair creation process of Fig. \ref{qedfig}b,
$\gamma+A
\to e^+e^- +A$, the scattering occurs off the {\em neutral} $e^+e^-$ pair. The
amplitude is now proportional to the dipole moment $\propto 1/m_e$ of the
pair, which
introduces a factor $\ktr/m_e$. The integral over momentum transfers is
logarithmic, $\sigma \propto \int d^2 \vec
\ktr/\ktr^2$, and has support from an extended region,
\beq
\mbox{$\ktr$ is hard} \Longleftrightarrow \frac{1}{R} \ll \ktr \ll m_e
\label{khard}
\eeq
The electron mass sets the upper limit on the logarithmic integral. Indeed,
for $\ktr \gg m_e$ the virtual electron or positron in Fig. \ref{qedfig}b is
far off-shell, and the fermion propagator causes the integral to converge.
\begin{figure}[htb]
\centerline{\vbox{\epsfxsize=12truecm\epsfbox{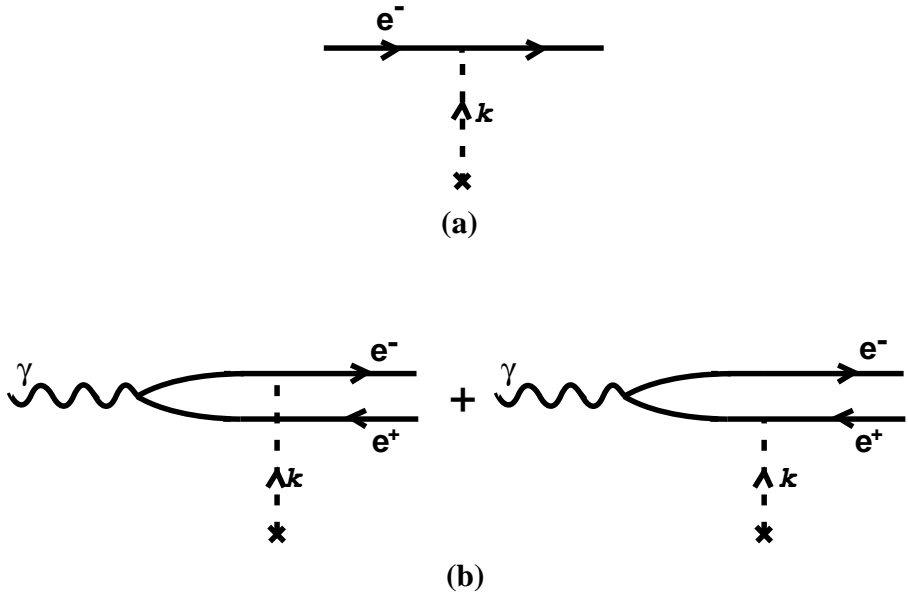}}}
\caption[*]{(a) Rutherford scattering of a charged particle off a nucleus.
(b) Pair creation process $\gamma+A \to e^+e^- +A$.}
\label{qedfig}
\end{figure}

The general classification of scattering into soft (monopole) and hard (dipole)
exchange is not altered by effects due to a running of the coupling
constant. In
the following QCD analysis we shall thus for clarity assume a fixed coupling
$\alpha_s$.

In the QCD process of Fig. \ref{genfig}, the incoming gluon has color charge,
while the final charmonium is a color singlet. In analogy to the QED case
above we may expect that the gluon will multiply
Rutherford scatter, with the
infrared cutoff provided by the inverse nucleon radius
$R^{-1} \sim \lqcd$.
At asymptotically large gluon energies the virtual $\cpair$ pair travels a
long distance, and it becomes highly probable that the pair is created far
upstream of the target. Hence the multiple scattering in the target will
actually occur off the quark pair rather than off the gluon. The compact pair
is, however, created in a color octet state and it
will behave like a gluon in soft rescattering processes.

On the other hand, it is intuitively plausible
that a scattering which changes the color {\it state} of the pair
(from singlet to octet or {\em vice versa}) will depend on the
differing structure of quark pairs and gluons,
\ie, such a scattering will probe the dipole moment of the pair.
As a consequence this scattering must be
hard\footnote{Conversely, a hard scattering may or may not
change the $\qpair$ color state. If the pair is turned into
a singlet then
any next scattering must also be hard, which implies a
subleading contribution in the multiple scattering production process we are
considering.}.
This applies, in particular, to the last scattering in Fig. \ref{genfig},
which by definition changes the color state of the pair from octet to
singlet.

In this paper we shall show explicitly how the above expectations are
realized in QCD, and also find the answer to the questions mentioned in the
beginning.

\begin{itemize}
\item[(i)]
Even though charge conjugation symmetry would
allow the $\jpsi$ to be produced through one hard and
one soft scattering in the target (without gluon emission), this would break
the factorization between hard and soft processes. The soft exchange
cannot be reliably calculated and such a contribution would in fact make it
impossible to predict $\jpsi$ production in PQCD.
We show that this problem does not arise since two {\it hard} scat\-terings are
needed to produce a $\jpsi$.

\item[(ii)]
We shall also see that the  broadening of $\langle \ptr^2 \rangle$ in nuclei
does in fact depend on the hard scale, as suggested by the data of
\eq{ptadep}. Multiplying the differential
cross section by a factor $\ptr^2$
makes the integral sensitive to the upper cut-off, and
$\langle \ptr^2 \rangle$ then depends (logarithmically) on the quark mass.
\end{itemize}

\mysection{$\eta_c$ production}

\subsection{Single gluon exchange}

The $^1$S$_0$ $\cpair$ state $\eta_c$ has positive charge conjugation and
thus couples to two gluons -- the projectile and one Coulomb gluon from the
target as shown in Fig. \ref{etac1gfig}.


\begin{figure}[htb]
\centerline{\vbox{\epsfxsize=12truecm\epsfbox{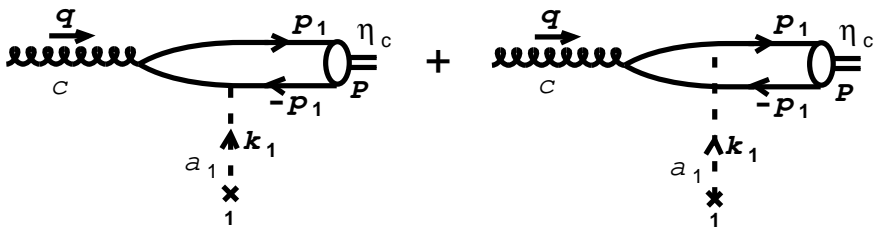}}}
\caption[*]{Amplitude for $\eta_c$ production induced by
one scattering.}
\label{etac1gfig}
\end{figure}

The amplitude for scattering from a fixed centre at position
$\xvec _1$ is
\beqa
\M(gg\to \eta_c) &=& -\frac{R_0g^3
}{\sqrt{4\pi m}}
 \frac{T^{a_1}_{A_1'A_1}}{\sqrt 3}
\int\frac{d^3\kvec_1}{(2\pi)^3} (2\pi)^3 \delta^3(\qvec+\kvec_1-2\pvec_1)
 \frac{\exp(-i\kvec_1\cdot\xvec_1)}{-\kvec_1^2}  \nonumber \\
&\times& \left({\cal A}_1+{\cal A}_2\right)
\nonumber \\
{\cal A}_1 &=& \tr(T^cT^{a_1}) \frac{\tr\left[\epsla(-\psla_1+\ksla_1+m)
\gamma^0 {{\cal P}_{S=0}} \right]}{(p_1-k_1)^2-m^2+\ieps}
\nonumber \\
{\cal A}_2 &=& \tr(T^cT^{a_1})
\frac{\tr\left[\gamma^0(\psla_1-\ksla_1+m)\epsla {{\cal P}_{S=0}}
\right]}{(p_1-k_1)^2-m^2+\ieps}
\label{etac1gamp}
\eeqa
Here the fixed centres are treated as heavy quarks of initial and final
color charge $A_1,A_1'$.
In the nonrelativistic limit of the
charmonium bound state the momenta of the $c$ and $\bar c$ are the same,
$p_1=p_2$, hence
\beq
q+k=2p_1 \equiv P  \label{momcons}
\eeq
where $k$ is the total momentum transfer to the target ($k=k_1$ in Fig.
\ref{etac1gfig}). The charm quark mass is denoted by $m$ and we have used the
operator \cite{projop}
\beq
{\cal P}_{S=0} = \frac{1}{\sqrt{2}}\gamma_5 (\psla_1+m) \label{proj0}
\eeq
to project the $\cpair$ pair onto the $\eta_c$ wave function at the origin,
which is given by $R_0$. The incoming gluon polarization vector is denoted
by $\varepsilon$.

The denominator in \eq{etac1gamp} is
\beq
(p_1-k_1)^2-m^2=(q-p_1)^2-m^2 = -2p_1\cdot q= -q\cdot k\label{den1}
\eeq
In the limit $\qo \to\infty$
and allowing the incoming gluon to have a finite
transverse momentum, \ie, $q \simeq (q^0, \qvec_{\perp},
q^0-q_{\perp}^2/2q^0)$,
we can write
\beqa
(p_1-k_1)^2-m^2 &\simeq& -2m_\perp^2  \label{den2} \\
\kz &\simeq& -\frac{2m_\perp^2 + \qvec_\perp \cdot \kvec_\perp}{\qo}
\label{kzexpr} \\
m_\perp^2 &\equiv& m^2 + k_{\perp}^2 /4  \label{mperpdef}
\eeqa
and obtain
an expression for the $\eta_c$ production amplitude whose
kinematics depends only on $\kvec_1$ (since $\kvec = \kvec_1$ here),
\beq
\M(g(q)g(k_1)\to \eta_c(2p_1)) \simeq
i\qo\frac{R_0g^3}{m_\perp^2\sqrt{24\pi m}}
 T^c_{A_1'A_1} \frac{\exp(-i\kvec_{1\perp}\cdot\xvec_{1\perp})}{k_{1\perp}^2}
\varepsilon_{0\mu\nu3}\,\varepsilon^\mu\,k_1^\nu  \label{ggetac}
\eeq
where $\varepsilon_{\mu\nu\rho\sigma}$ is the fully antisymmetric
Levi-Civita tensor.
In the following, the direction of the incoming gluon will be chosen
along the $z$-axis, \ie, we take $\qvec_\perp = \vec{0}$ from now on.

We can now make the following observations.
\begin{itemize}
\item The $\eta_c$ production amplitude $\M(gg\to \eta_c)$ is proportional
to $\kvec_{1\perp}/k_{1\perp}^2$, \ie, the effective values of $k_{1\perp}$
are hard,
\beq
\lqcd \ll k_{1\perp} \ll m  \label{k1range}
\eeq
which justifies the use of perturbative QCD for this process.
\item The fact that the effective upper limit of $k_{1\perp}$ is given by
$m$ originates from the quark propagator, proportional to $1/m_\perp^2 =
1/(m^2+k_{1\perp}^2/4)$.
As we shall see and as already announced in Eqs. (\ref{ksoft}) and
(\ref{khard}), the $\kvec_\perp$ dependence of $m_\perp$ may be safely
neglected when calculating amplitudes or cross sections.
However, this dependence can
be important in the calculation of
other quantities, like $\ave{p_\perp^2}$ for $\eta_c$ or $\jpsi$, since
weighting the differential cross section by $p_\perp^2$ can shift the
effective values of the transverse momenta to be of order $m$.
For this reason we will keep the $m_\perp$ dependence in the following.
\item In this process, the incoming $\cpair$ pair is in a color octet state,
but $k_{1\perp}$ is not soft as it would have been in the case of elastic
gluon scattering. In order to turn the color octet $\cpair$ into a singlet the
exchanged gluon has to probe the color dipole moment of the pair.
\end{itemize}

In the next subsection we consider the case when the $\cpair$ pair
undergoes {\em two} Coulomb scatterings in the target.

\subsection{$\eta_c$ production through two gluon exchange}

According to the notation of Fig. \ref{genfig} we assume momentum transfers
$\kvec_1,\ \kvec_2$ from the two scattering centres located at $\xvec_1,\
\xvec_2$, respectively. The longitudinal coordinates of the centres are taken
to be $x_1^z=0,\ x_2^z=\tau>0$ (we do not consider double scattering from
a single centre in this paper).
The calculation, which is summarized below,
shows the following.

\begin{itemize}
\item $k_{1\perp}$ is soft, whereas $k_{2\perp}$ is hard (in the sense
of Eqs. (\ref{ksoft}) and (\ref{khard})), \ie,
\beq
k_{1\perp} \sim\lqcd;\ \ \ \ \ \lqcd\ll k_{2\perp} \ll m \label{kingg}
\eeq
\item The amplitude {\em factorizes}. It is given by a convolution of the
incoming gluon elastic scattering amplitude $gg\to g$ and the $\eta_c$
production amplitude $gg\to \eta_c$ given by \eq{ggetac},
\beqa
i\M(ggg\to \eta_c) &=& \sum_{\lambda'}\int\frac{d^2\kvec_{1\perp}}{(2\pi)^2}
i\M_{el}[g(q,\lambda)g(k_1) \to g(q+k_1,\lambda')]\, \frac{1}{2\qo}
\nonumber \\ &\times&
i\M[g(q+k_1,\lambda')g(k_2) \to \eta_c(q+k)] \label{eta2conv}
\eeqa
where $\lambda'$ is the polarization of the intermediate gluon and
\beqa
i\M_{el}[g(q,\lambda)g(k_1) \to g(q+k_1,\lambda')] &=& 2 q^0 g^2
\varepsilon(\lambda)\cdot\varepsilon(\lambda')
f_{ca_{1}c'} T^{a_1}_{A_1'A_1}  \nonumber \\
&\times& \frac{\exp(-i\kvec_{1\perp}\cdot\xvec_{1\perp})}{k_{1\perp}^2}
\label{ggel}
\eeqa
This equation may be diagrammatically represented as
\beq
\mbox{\raisebox{-1.1truecm}{\vbox{\epsfxsize=3.5truecm\epsfbox{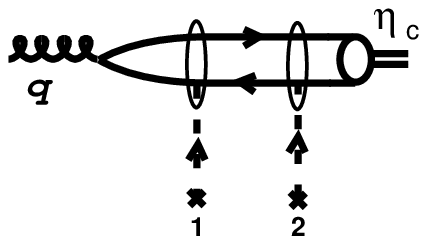}}}}
= \sum_{\lambda'}\int\frac{d^2\kvec_{1\perp}}{(2\pi)^2}
\mbox{\raisebox{-1truecm}{\vbox{\epsfxsize=2truecm\epsfbox{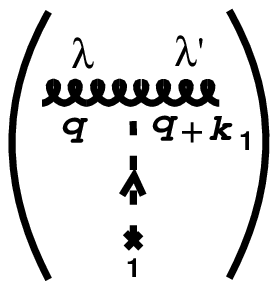}}}}
\,\frac{1}{2\qo}
\mbox{\raisebox{-1cm}{\vbox{\epsfxsize=3truecm\epsfbox{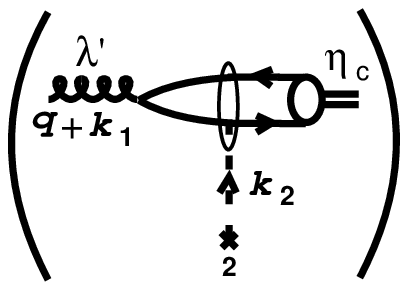}}}}
\label{figeq1}
\eeq
\end{itemize}
Let us now present the derivation of \eq{eta2conv}, which within our
approach is exact in the limit $\qo\to\infty$. We write the amplitude as
\beqa
\M(ggg\to \eta_c) &=& -\frac{R_0g^5}{\sqrt{24\pi
m}} T^{a_1}_{A_1'A_1} T^{a_2}_{A_2'A_2}
\int\frac{d^3\kvec_1}{(2\pi)^3} \exp[i\tau(\kz_1-\kz)]  \nonumber \\
&\times&
\frac{\exp(-i\kvec_{1\perp}\cdot
\xvec_{1\perp} -i\kvec_{2\perp}\cdot \xvec_{2\perp})}{k_{1\perp}^2 \
k_{2\perp}^2}   ({\cal A+B})
\label{gggetac}
\eeqa
where the amplitudes ${\cal A}$ (${\cal B}$) correspond to the 4 (8)
diagrams shown in Fig. \ref{gggetacfig}a (b).


\begin{figure}[hp]
\centerline{\vbox{\epsfxsize=16truecm\epsfbox{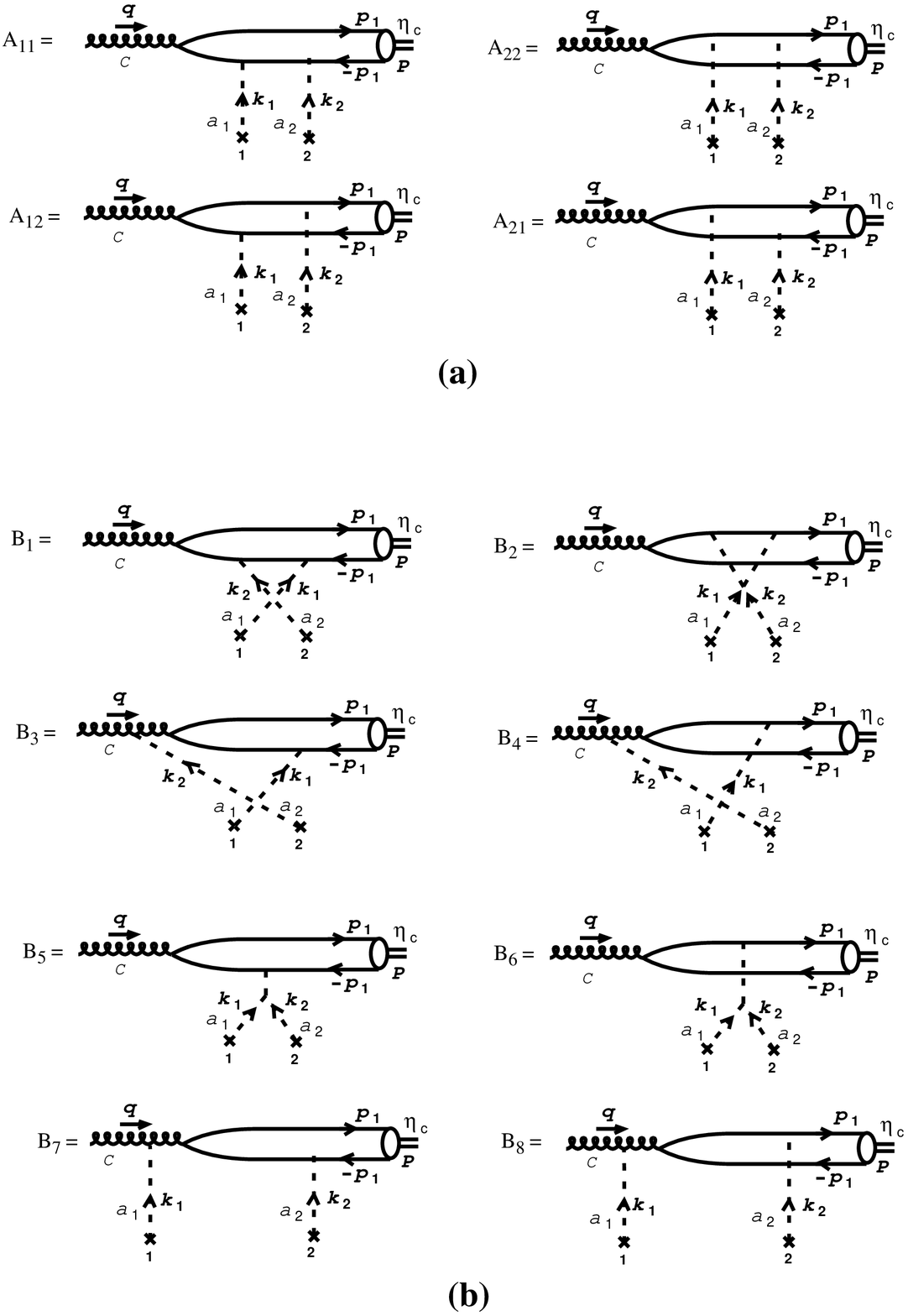}}\hspace{2cm}}
\vspace{2ex}
\caption[*]{Diagrams for the $ggg\to \eta_c$ production amplitude.}
\label{gggetacfig}
\end{figure}

Since we take $\tau>0$ the
integral over $k_1^z$ can be evaluated by closing the integration contour in
the upper half plane.

It is rather straightforward to see that none of the diagrams of Fig.
\ref{gggetacfig}b contribute in our limit, \ie, ${\cal B}=0$. Namely,
\begin{itemize}
\item[--] Since the Coulomb exchange is instantaneous, the diagrams $B_1
\ldots B_4$ that have crossed exchanges involve the creation of $\qpair$
pairs with energies $\propto \qo\to\infty$. All these diagrams have quark
propagators whose residues in the Im\,$\kz_1>0$ plane vanish asymptotically.

\item[--] Diagrams $B_5$ and $B_6$ which involve the three-gluon vertex are
also negligible. Our static scattering centres and the $\qo\to\infty$ limit
restrict the Lorentz indices to be $\mu=\nu=0$
for the two gluons attached to the centres and $\sigma=0,3$
for the gluon attached to the $\qpair$. Hence the vertex is
at most on the order of the asymptotically vanishing
longitudinal momentum transfers
(\cf\ Eqs. (\ref{kzexpr}) and (\ref{kzpole}) below).

\item[--] In diagrams $B_7$ and $B_8$ the first
rescattering gluon is attached to
the projectile gluon. In the $\qo\to\infty$ limit
the $g\to\qpair$ fluctuation will occur long before the target, and thus
such scattering should be suppressed.
Technically this appears in the following way.
The product of the gluon and quark propagators
in diagrams  $B_7$ and $B_8$ involves the denominator
\beqa
[(q+k_1)^2+\ieps]\,[(p_1-k_2)^2-m^2+\ieps] \simeq \nonumber \\
2(\qo)^2\,[\kz_1+\frac{\kvec_{1\perp}^2}{2\qo}-\ieps]\,
[\kz_1-\kz-\frac{\kvec_{1\perp}\cdot\kvec_{2\perp}}{\qo}-\ieps]\ \>,
\nonumber
\eeqa
and thus has two poles in the upper  $\kz_1$ half-plane,
whose residues can easily be seen to cancel in the  $\qo\to\infty$ limit.
\end{itemize}

Thus, only the four diagrams of Fig. \ref{gggetacfig}a are nonvanishing and
contribute to the ${\cal A}$-term in \eq{gggetac}. Separating the color
factors $c_{ij}$, the Lorentz traces $\tr_{ij}$ and the denominators
$\Delta_{ij}$ we write
\beq
{\cal A}= \sum_{i,j} {A}_{ij} = \sum_{i,j} c_{ij}
\frac{\tr_{ij}}{\Delta_{ij}}
\label{asep}
\eeq
where
\beqa
c_{11} &=& c_{12} = \tr(T^cT^{a_1}T^{a_2})  \nonumber \\
c_{22} &=& c_{21} = \tr(T^cT^{a_2}T^{a_1})  \label{cij}
\eeqa
\beqa
\tr_{11} &=& \tr\left[\epsla(-\psla_1+\ksla_1+\ksla_2+m)\gamma^0
(-\psla_1+\ksla_2+m) \gamma^0(\psla_1-m)\gamma_5 \right]  \nonumber \\
\tr_{22} &=& \tr\left[\gamma^0(\psla_1-\ksla_2+m)\gamma^0
(\psla_1-\ksla_1-\ksla_2+m)\epsla (\psla_1-m)\gamma_5 \right]  \nonumber \\
\tr_{12} &=& \tr\left[\gamma^0(\psla_1-\ksla_2+m)\epsla (-\psla_1+\ksla_1+m)
\gamma^0 (\psla_1-m)\gamma_5 \right]  \nonumber \\
\tr_{21} &=& \tr\left[\gamma^0(\psla_1-\ksla_1+m)\epsla (-\psla_1+\ksla_2+m)
\gamma^0 (\psla_1-m)\gamma_5 \right]  \label{trij}
\eeqa
\beqa
\Delta_{11} &=& \Delta_{22} = [(p_1-k)^2-m^2+\ieps] [(p_1-k_2)^2-m^2+\ieps]
 \nonumber \\
\Delta_{12} &=& \Delta_{21} = [(p_1-k_1)^2-m^2+\ieps] [(p_1-k_2)^2-m^2+\ieps]
\label{deltaij}
\eeqa

Reversing the order of the $\gamma$-matrices in $\tr_{22}$ and $\tr_{21}$ one
easily finds
\beq
\tr_{11}=-\tr_{22}\ ;\ \ \ \ \tr_{12}=-\tr_{21}  \label{trrel}
\eeq
In QED \eq{trrel} ensures a vanishing transition amplitude between a
three-photon state (of negative charge conjugation, $C=-$) and a $^1$S$_0$
positronium state $(C=+)$. Due to the color charges of QCD the analogous
amplitude does not vanish but is proportional to the fully antisymmetric
structure constants $f_{abc}$ of SU(3),
\beqa
{\cal A} &=& \left[\tr(T^cT^{a_1}T^{a_2})-\tr(T^cT^{a_2}T^{a_1}) \right]
\left[ \frac{\tr_{11}}{\Delta_{11}} + \frac{\tr_{12}}{\Delta_{12}} \right]
\nonumber \\
&=& \frac{i}{2} f_{ca_1a_2}
\left[ \frac{\tr_{11}}{\Delta_{11}} + \frac{\tr_{12}}{\Delta_{12}} \right]
\label{fabcexpr}
\eeqa
An evaluation of the traces gives, in the $q^0 \to \infty$ limit,
\beqa
\tr_{11} &=& -2i(\qo)^2\,\varepsilon_{0\mu\nu3}\,\varepsilon^\mu\,(k_1+k_2)^\nu
\nonumber \\
\tr_{12} &=&
-2i(\qo)^2\,\varepsilon_{0\mu\nu3}\,\varepsilon^\mu\,(-k_1+k_2)^\nu
\label{trexpr} \ \>.
\eeqa
The denominators may be simplified using
\beqa
(p_1-k_1)^2-m^2+\ieps &\simeq& \qo\,(\kz_1+\frac{\kvec_{1\perp} \cdot
\kvec_{2\perp}}{\qo}+\ieps)  \nonumber \\
(p_1-k_2)^2-m^2+\ieps &\simeq& -\qo\,(\kz_1-\kz-\frac{\kvec_{1\perp} \cdot
\kvec_{2\perp}}{\qo}-\ieps) \>.
\label{denexpr}
\eeqa

Integrating over $\kz_1$ in \eq{gggetac} by closing the contour in the upper
half-plane picks according to \eq{denexpr} the pole
\beq
\kz_1=\kz + \frac{\kvec_{1\perp} \cdot \kvec_{2\perp}}{\qo} +\ieps
\label{kzpole}
\eeq
such that
\beqa
\int\frac{d\kz_1}{2\pi}\,{\cal A}\,\exp[i\tau(\kz_1-\kz)] &=&
i(\qo)^2 f_{ca_1a_2} \varepsilon_{0\mu\nu3}\varepsilon^\mu \nonumber \\
&\times&\left[\frac{(k_1+k_2)^\nu}{\qo\kz(-\qo)} +
\frac{(-k_1+k_2)^\nu}{(-\qo)(\qo\kz+2\kvec_{1\perp}\cdot\kvec_{2\perp})}
\right] \hspace{1cm}
\label{inta}
\eeqa
Note that the poles of the Coulomb gluon propagators lead to a subleading
contribution in the $q_0 \to \infty$ limit.
Using \eq{kzexpr} for $\qvec_\perp = \vec{0}$ and
\beq
|\kvec_{1\perp}\cdot\kvec_{2\perp}| \ll m_\perp^2
\label{etacvalid}
\eeq
we find for the
full amplitude of \eq{gggetac}
\beqa
\M(ggg\to \eta_c) &=& -i\qo\frac{R_0g^5}{m_\perp^2
\sqrt{24\pi m}} T^{a_1}_{A_1'A_1} T^{a_2}_{A_2'A_2}
f_{ca_1a_2}\,\varepsilon_{0\mu\nu3}\,\varepsilon^\mu \nonumber \\
&\times&\int\frac{d^2\kvec_{1\perp}}{(2\pi)^2}
\exp(-i\kvec_{1\perp}\cdot \xvec_{1\perp} -i\kvec_{2\perp}\cdot
\xvec_{2\perp}) \frac{1}{k_{1\perp}^2} \frac{k_2^\nu}{k_{2\perp}^2}
\label{gggetacfin}
\eeqa
Writing
\beq
\varepsilon^\mu(\lambda) = -\sum_{\lambda'} \left[\varepsilon(\lambda)
\cdot \varepsilon(\lambda')^*\right] \varepsilon^\mu(\lambda') \label{epscom}
\eeq
directly leads to the factorized expression of \eq{eta2conv}.

Let us note that \eq{gggetacfin} involves the denominator
$m_\perp^2 = m^2 + (\kvec_{1\perp}+\kvec_{2\perp})^{2}/4$,
instead of $(m^2 + \kvec_{2\perp}^{\ 2}/4)$ for the $\eta_c$
production amplitude contained in \eq{eta2conv}. By calculating
$\M (ggg \to \eta_c)$ as a function of $\kvec_\perp$ and
$\xvec_{2\perp}-\xvec_{1\perp}$, one shows that the
effective values of the transverse momenta satisfy $k_{1\perp} \ll
k_{2\perp} \sim k_{\perp}$ for all
$|\xvec_{2\perp}-\xvec_{1\perp}|$
(provided $k_\perp \gg
\Lambda_{QCD}$), so that \eq{eta2conv} follows.
However, as mentioned earlier the $\kvec_{1\perp}$ dependence of $m_\perp$
can be important when calculating $\ave{p_\perp^2}$, for which the correct
expression to start with will be \eq{gggetacfin}.

The
$\kvec_{2\perp}$ factor in \eq{gggetacfin} explicitly shows that the second
exchange is hard in the sense of \eq{khard}. The amplitude to turn the color
octet $\cpair$ pair into a color singlet is thus proportional to the dipole
moment of the pair.
In the next section we study the analogous situation in $\jpsi$ production,
which differs from the case of the $\eta_c$ since a minimum of two exchanges
with the target is now required.

\mysection{$\jpsi$ production}

\subsection{Two gluon exchange}

The $\jpsi$ is a $^3$S$_1$ charmonium state with negative charge conjugation,
and hence couples to a minimum of three gluons. The lowest order amplitude
(which does not involve gluons in the final state) is thus given by the
diagrams of Fig. \ref{gggetacfig}a, with the $\eta_c$ replaced by the
$\jpsi$. For the reasons discussed in the previous section the diagrams
$B_1$ and $B_2$
of Fig. \ref{gggetacfig}b do not contribute to high energy $\jpsi$
production. Hence the amplitude may again be written as in \eq{gggetac} with
${\cal B}=0$. In the expression for ${\cal A}$, the only difference is that
the projection operator (\ref{proj0}) is replaced by a projection onto a
vector state \cite{projop},
\beq
{\cal P}_{S=1} = -\frac{1}{\sqrt{2}}\esla (\psla_1+m) \label{proj1}
\eeq
where $e(S_z)$ is the $\jpsi$ polarization vector, defined in the $\jpsi$
rest frame as
\beqa
e(S_z=\pm 1) &\equiv& e_T= (0,1,\pm i,0)/\sqrt{2}\ \ \ \ \mbox{{\rm for
transverse }}\jpsi\   \nonumber \\
e(S_z=0) &\equiv& e_L = (0,\vec 0_\perp,1)
\ \ \ \ \ \ \ \ \ \ \ \ \mbox{{\rm   for longitudinal }}\jpsi\
\label{polvect}
\eeqa
We shall use the Gottfried-Jackson frame, where the $z$-axis in the $\jpsi$
rest frame is taken parallel to the projectile momentum $\qvec$. The
invariant product $e\cdot w$ for any vector $w$ can then be expressed,
in the high energy limit we consider, as
\beqa
e_T \cdot w &=& -\vec e_\perp \cdot \vec w_\perp  \nonumber \\
e_L \cdot w &=& m \frac{q\cdot w}{q\cdot p_1} - \frac{p_1\cdot w}{m}
\label{polprod}
\eeqa

From the trace expressions corresponding to \eq{trij} it can be seen that
for the $\jpsi$ amplitude
\beq
\tr_{11}=\tr_{22}\ ;\ \ \ \ \tr_{12}=\tr_{21}  \label{trreljpsi}
\eeq
which differ from the $\eta_c$ case of \eq{trrel} by a sign. It is in fact
easily seen that \eq{trreljpsi} for the $\jpsi$ generalizes in the case of
$n$ scatterings to
\beq
\tr_A = (-1)^n\, \tr_{A^*}   \label{trreljpsin}
\eeq
where $A$ and $A^*$ are charge conjugated diagrams. \eq{trreljpsin} is
readily understood as a consequence of charge conjugation invariance in
QED.

The expression for the amplitude ${\cal A}$ in \eq{gggetac} for $\jpsi$
production is then
\beqa
{\cal A} &=& \left[\tr(T^cT^{a_1}T^{a_2})+\tr(T^cT^{a_2}T^{a_1}) \right]
\left[ \frac{\tr_{11}}{\Delta_{11}} + \frac{\tr_{12}}{\Delta_{12}} \right]
\nonumber \\
&=& \frac{1}{2} d_{ca_1a_2}
\left[ \frac{\tr_{11}}{\Delta_{11}} + \frac{\tr_{12}}{\Delta_{12}} \right]
\label{dabcexpr}
\eeqa
In the limit $\qo\to\infty$ the explicit expressions for the traces are
\beq
\tr_{11}=-\tr_{12}=-4m(\qo)^2 \beta_{T,L}(\lambda)  \label{trjpsi}
\eeq
where $\beta$ depends on the polarization $(\lambda)$ of the incoming gluon
and on that of the $\jpsi$ $(T,L)$,
\beqa
\beta_T(\lambda) &=& -\varepsilon(\lambda) \cdot e_T  \nonumber \\
\beta_L(\lambda) &=& 2\varepsilon(\lambda) \cdot p_1 \,m/m_\perp^2
\label{betaexpr}
\eeqa
For the general case of $n$ scatterings (\cf\ Fig. \ref{genfig}) the trace
for a given diagram is
\beq
\tr = -4m (\qo)^n \beta_{T,L}(\lambda) (-1)^{n_{\bar q}}   \label{gentr}
\eeq
where $n_{\bar q}$ is the number of Coulomb gluons attached to the antiquark.

Using now Eqs. (\ref{dabcexpr}) and (\ref{trjpsi}) in the expression
(\ref{gggetac}) and performing the $\kz_1$-integral by picking up the pole
(\ref{kzpole}) we find (once again $|\kvec_{1\perp}\cdot\kvec_{2\perp}| \ll
m_\perp^2$),
\beqa
\M(g(\lambda)gg\to \jpsi) &=& -i\qo
m\frac{R_0g^5}{m_\perp^4
\sqrt{24\pi m}} T^{a_1}_{A_1'A_1} T^{a_2}_{A_2'A_2} d_{ca_1a_2}
\beta_{T,L}(\lambda) \hspace{1cm} \nonumber \\
&\times&\int\frac{d^2\kvec_{1\perp}}{(2\pi)^2}
\exp(-i\kvec_{1\perp}\cdot \xvec_{1\perp} -i\kvec_{2\perp}\cdot
\xvec_{2\perp})
\frac{\kvec_{1\perp} \cdot \kvec_{2\perp}}{k_{1\perp}^2\,k_{2\perp}^2}
\hspace{1cm}
\label{gggjpsifin}
\eeqa

Since the numerator of the $\jpsi$ production amplitude (\ref{gggjpsifin})
is proportional to $\kvec_{1\perp}$ and $\kvec_{2\perp}$ it follows that
both transferred momenta are hard in the sense of \eq{khard}. This
contribution to $\jpsi$ production is then a higher twist effect compared to
the standard one involving a single target scattering and a radiated gluon,
as is required by factorization between hard and soft processes in
PQCD.

Next we shall investigate the case of one `extra' target scattering. This
is of importance for understanding the general systematics of rescattering
effects in $\jpsi$ production and the ${p_\perp}$-broadening
effects in nuclei.

\subsection{Three gluon exchange in $\jpsi$ production}

In the high energy limit the $\jpsi$ production amplitude induced by three
scatterings on static centres is given by the eight diagrams
of Fig. \ref{gggjpsifig}.

\begin{figure}[b]
\centerline{\vbox{\epsfxsize=10truecm\epsfbox{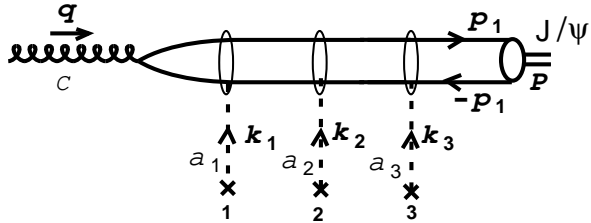}}\hspace{2cm}}
\caption[*]{$\jpsi$ production amplitude induced by three scatterings.
Eight diagrams are generated by attaching each Coulomb gluon to one of
the quark lines.}
\label{gggjpsifig}
\end{figure}

From the $\eta_c$ result (\ref{gggetacfin}) we may expect that the
last transfer $(k_3)$, which turns the $\cpair$ pair into a color singlet,
will be hard. It is less obvious if either one of the first two exchanges can
be soft. As already noticed, the propagating quark pair stays
dominantly in a color octet state
even when undergoing a hard scattering (except for the last one).
Therefore, one would expect it to suffer soft rescattering both
before and after the first hard exchange. This is indeed what we shall find
below.

Taking $x_1^z=0,\ x_2^z=\tau$ and $x_3^z=\tau'>\tau>0$ the amplitude can be
expressed as
\beqa
\M &=& -\frac{R_0g^7}{\sqrt{24\pi m}}
\int\frac{d^3\kvec_1}{(2\pi)^3} \frac{d^3\kvec_2}{(2\pi)^3}
\prod_{j=1}^3 \left[ T^{a_j}_{A_j'A_j}
\frac{\exp(-i\kvec_{j\perp}\cdot \xvec_{j\perp})}{k_{j\perp}^2} \right]
\nonumber\\
&\times& \exp \left[ -i\kz\tau' +i\kz_1\tau'  + i\kz_2(\tau'-\tau) \right]
{\cal A}
\label{4gjpsi}
\eeqa
After some algebra we find, with an obvious notation for color traces,
\beqa
{\cal A}&=&4m(\qo)^3 \beta_{T,L}(\lambda) \left\{ \left[\tr(ca_1a_2a_3) -
\tr(a_1ca_3a_2)
\right] \left(\frac{1}{\Delta_1}-\frac{1}{\Delta_5} \right) \right.
\nonumber \\
&+& \left. \left[\tr(a_1ca_2a_3) - \tr(ca_1a_3a_2)
\right] \left(\frac{1}{\Delta_7}-\frac{1}{\Delta_3} \right) \right\}
\label{4ga}
\eeqa
where
\beqa
\Delta_1 &=& [(p_1-q)^2-m^2+\ieps] [(p_1-k_2-k_3)^2-m^2+\ieps]
[(p_1-k_3)^2-m^2+\ieps]  \nonumber \\
\Delta_5 &=& [(p_1-k_1-k_2)^2-m^2+\ieps] [(p_1-k_2)^2-m^2+\ieps]
[(p_1-k_3)^2-m^2+\ieps]  \nonumber \\
\Delta_3 &=& [(p_1-k_1)^2-m^2+\ieps] [(p_1-k_2-k_3)^2-m^2+\ieps]
[(p_1-k_3)^2-m^2+\ieps]  \nonumber \\
\Delta_7 &=& [(p_1-k_2)^2-m^2+\ieps] [(p_1-k_1-k_3)^2-m^2+\ieps]
[(p_1-k_3)^2-m^2+\ieps]  \nonumber \\ \label{4gden}
\eeqa

In the $\qo\to\infty$ limit we have
\beqa
(p_1-q)^2-m^2 &\simeq& -2m_\perp^2  \nonumber \\
(p_1-k_i)^2-m^2+\ieps &\simeq& \qo[\kz_i+ \kvec_{i\perp} \cdot
(\kvec_{j\perp}+\kvec_{k\perp})/\qo +\ieps] \nonumber \\
(p_1-k_i-k_j)^2-m^2+\ieps &\simeq& \qo[\kz_i+\kz_j+ \kvec_{k\perp} \cdot
(\kvec_{i\perp}+\kvec_{j\perp})/\qo +\ieps] \label{4gappr}
\eeqa
where $i,j,k$ is any cyclic combination of $1,2,3$. Integrating over $\kz_2$
and then over $\kz_1$ picks the poles
\beq
\kz_2=\kz-\kz_1+ (\kvec_{1\perp}+\kvec_{2\perp})\cdot \kvec_{3\perp}/\qo
+\ieps  \label{4gk2pole}
\eeq
and
\beqa
\kz_1&=&\kz +\kvec_{1\perp} \cdot (\kvec_{2\perp}+\kvec_{3\perp})/\qo +\ieps
\nonumber \\
\kz_1&=&\kz +\kvec_{1\perp} \cdot (\kvec_{2\perp}+\kvec_{3\perp})/\qo +
2\kvec_{2\perp} \cdot \kvec_{3\perp}/\qo + \ieps  \label{4gk1pole}
\eeqa
As in the cases previously studied, the {\em first longitudinal transfer}
$\kz_1 \simeq \kz$ puts the $\cpair$ pair on-shell, while the succeeding
transfers $\kz_i \ll \kz$ maintain this on-shellness. This is because
\beq
|\kvec_{i\perp}\cdot \kvec_{j\perp}| \ll m_\perp^2  \label{kijrel}
\eeq
holds\footnote{Note that $|\kvec_{i\perp}\cdot \kvec_{j\perp}| \ll m_\perp^2$
also holds
when one transverse momentum $|\kvec_{i\perp}|$ is
of order $m$ and the two others are
smaller:  $|\kvec_{j\perp}| \ll m$
for $j \neq i$. We have checked that this is indeed true in the
calculation of $\ave{p_\perp^2}$ for transverse and also longitudinal
$\jpsi$
in the logarithmic approximation $\log({m^2}/{\Lambda_{QCD}^2}) \gg 1$ .}
in both relevant ranges (\ref{ksoft}) and (\ref{khard}). We finally
obtain
\beqa
\M(4g\to \jpsi) &=&
im\qo\frac{R_0g^7}{m_\perp^4\sqrt{24\pi m}} \beta_{T,L}(\lambda)
\hspace{4cm} \nonumber \\ &\times&
\int\frac{d^2\kvec_1}{(2\pi)^2} \frac{d^2\kvec_2}{(2\pi)^2}
\prod_{j=1}^3 \left[ T^{a_j}_{A_j'A_j} \,
\frac{\exp(-i\kvec_{j\perp}\cdot \xvec_{j\perp})}{k_{j\perp}^2} \right]
\nonumber\\
&\times& \left(
d_{ca_1d} f_{a_2a_3d}\, \kvec_{1\perp}\cdot \kvec_{3\perp} +
f_{ca_1d} d_{a_2a_3d}\, \kvec_{2\perp}\cdot \kvec_{3\perp} \right)
\label{4gjpsifin}
\eeqa

 From \eq{4gjpsifin} we deduce the following:
\begin{itemize}
\item The last momentum transfer $k_3$ is hard.
\item One of the first two transfers is hard and the other is soft.
\end{itemize}
Let us consider the two parts of the amplitude (\ref{4gjpsifin}) separately,
\beq
\M = \M_1 + \M_2  \label{mdecomp}
\eeq
where the index indicates which is the first hard transfer. It is
straightforward to
see that the $\M_2$ part factorizes in a way
analogous\footnote{The difference in the expressions of $m_\perp^2$
in \eq{4gjpsifin} and in the $\jpsi$ production amplitude appearing
in \eq{m2conv}, respectively
$m^2 + (\kvec_{1\perp}+\kvec_{2\perp}+\kvec_{3\perp})^{2}/4$
and
$m^2 + (\kvec_{2\perp}+\kvec_{3\perp})^{2}/4$,
is negligible. See the comments for the case of $\eta_c$ in the end
of section 2.}
to \eq{eta2conv},
\beqa
i\M_2 &=& \sum_{\lambda'}\int\frac{d^2\kvec_{1\perp}}{(2\pi)^2}
i\M_{el}[g(q,\lambda)g(k_1) \to g(q+k_1,\lambda')]\, \frac{1}{2\qo}
\nonumber \\ &\times&
i\M[g(q+k_1,\lambda')g(k_2)g(k_3) \to \jpsi(q+k)]
\label{m2conv}
\eeqa
or in pictorial form
\beq
\mbox{\raisebox{-1.1truecm}{\vbox{\epsfxsize=4truecm\epsfbox{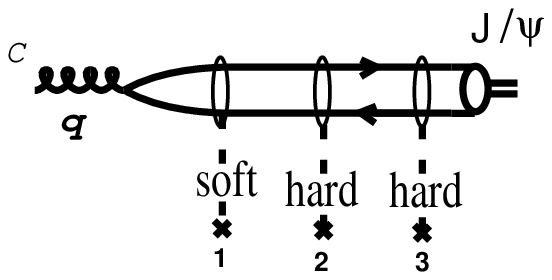}}}}
= \sum_{\lambda'}\int\frac{d^2\kvec_{1\perp}}{(2\pi)^2}
\mbox{\raisebox{-.9truecm}{\vbox{\epsfxsize=2truecm\epsfbox{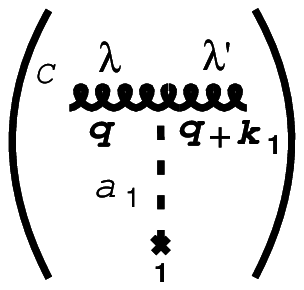}}}}
\,\frac{1}{2\qo}
\mbox{\raisebox{-1truecm}{\vbox{\epsfxsize=4truecm\epsfbox{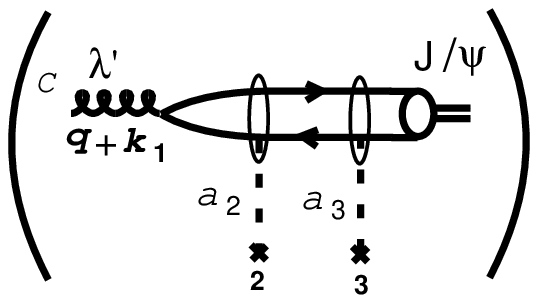}}}}
\label{figeq2}
\eeq
The soft transfer $k_{1\perp}$ thus mimics rescattering of the initial gluon.
(As remarked above, however, the longitudinal part $\kz_1$ is
relatively big in the sense that it puts the incoming heavy quark pair on its
mass-shell, as seen from \eq{4gk1pole}.)

Factorization is somewhat more involved for the $\M_1$ term of \eq{mdecomp},
where $k_{1\perp}$ is hard. A relation like
\beq
\mbox{\raisebox{-1.1truecm}{\vbox{\epsfxsize=4truecm\epsfbox{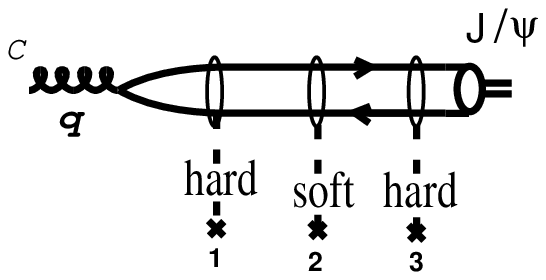}}}}
= \sum_{\lambda'}\int
\frac{d^2\kvec_{2\perp}}{(2\pi)^2}
\mbox{\raisebox{-.9truecm}{\vbox{\epsfxsize=2truecm\epsfbox{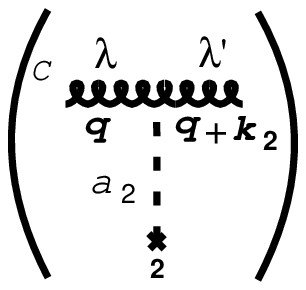}}}}
\,\frac{1}{2\qo}
\mbox{\raisebox{-1truecm}{\vbox{\epsfxsize=4truecm\epsfbox{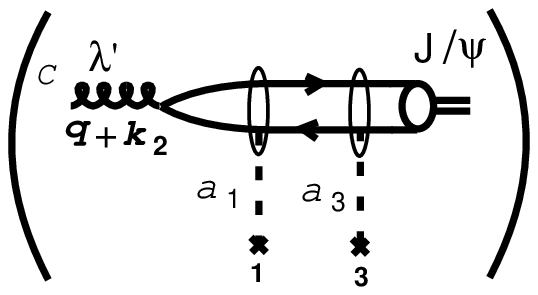}}}}
\label{figeq3}
\eeq
cannot hold, since the soft $k_{2\perp}$ exchange sees a color which has been
`rotated' away from the incoming color $c$ by the $k_1$ gluon. Rather, the
color structure of the $\M_1$ term in \eq{4gjpsifin} can be pictorially
represented as
\beqa
&\ & \mbox{\hspace{3.6truecm}$(i\M_1)_{color} = d_{ca_1d}\,if_{a_2a_3d}$}
\nonumber \\
\nonumber \\
&\ & \mbox{$=\ \mbox{\raisebox{-1truecm}{\vbox{\epsfxsize=11truecm
\epsfbox{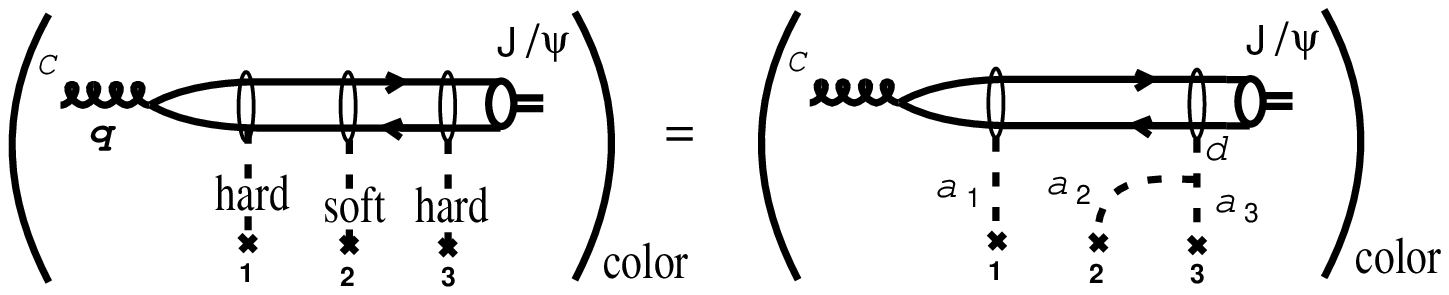}}}}$}
\label{m1color}
\eeqa
On the other hand, forgetting the color factors and considering only the
Lorentz structure of $\M_1$, we easily see from \eq{4gjpsifin} the validity
of the factorization formula
\beqa
&\ & \mbox{\hspace{1.5truecm}$(i\M_1)_{Lorentz} =
\mbox{\raisebox{-1truecm}{\vbox{\epsfxsize=5.5truecm\epsfbox{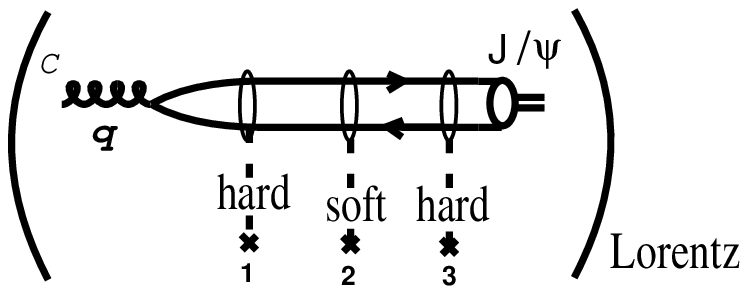}}}}$}
\nonumber \\
\nonumber \\
&\ & = \sum_{\lambda'}\int
\frac{d^2\kvec_{2\perp}}{(2\pi)^2}
\ \ \ \
\mbox{\raisebox{-1truecm}{\vbox{\epsfxsize=3truecm\epsfbox{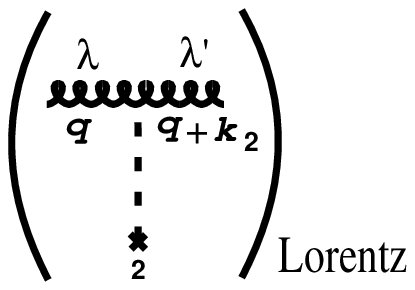}}}}
\,\frac{1}{2\qo}\ \ \ \
\mbox{\raisebox{-1.1truecm}{\vbox{\epsfxsize=4.5truecm\epsfbox{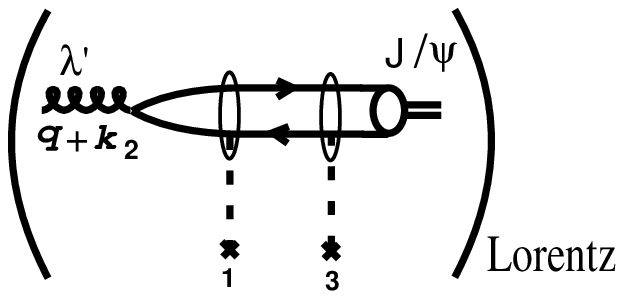}}}}
\label{m1lorentz}
\eeqa
The differing structure of Eqs. (\ref{m1color}) and (\ref{m1lorentz})
prevents us from factorizing the $\M_1$ part of the amplitude in a
straightforward way analogous to \eq{m2conv} for $\M_2$. The soft gluon
$(k_2)$ between the hard transfers $(k_1,k_3)$ changes the color structure
of the hard production amplitude and, as we shall see below, ensures that
there is no interference in the cross section between $\M_1$ and $\M_2$.
It is convenient to
summarize our result for the full amplitude (\ref{4gjpsifin}) in the form
\beqa
\M(4g\to \jpsi) &=& \M_1+\M_2  \nonumber \\
&=& \prod_{j=1}^3 \left[ T^{a_j}_{A_j'A_j} \right]
\left( if_{a_2a_3d}d_{ca_1d}\widetilde\M_1 +
if_{ca_1d} d_{a_2a_3d}\widetilde\M_2 \right) \label{4gjpsifact}
\eeqa
where\footnote{We leave for convenience the $1 \over \sqrt 3$ color factor
for the bound state in the Lorentz part.}
$\widetilde\M_i= (\M_i)_{Lorentz}$.
For instance, \eq{m1lorentz} is explicitly
\beqa
i\widetilde\M_1 &=& \sum_{\lambda'}\int\frac{d^2\kvec_{2\perp}}{(2\pi)^2}
i\widetilde\M_{el}[g(q,\lambda)g(k_2) \to g(q+k_2,\lambda')]\,
\frac{1}{2\qo}
\nonumber \\ &\times&
i\widetilde\M[g(q+k_2,\lambda')g(k_1)g(k_3) \to \jpsi(q+k)]
\label{mtildeconv}
\eeqa
where
\beqa
\widetilde\M_{el}[g(q,\lambda)g(k_2) \to g(q+k_2,\lambda')] \hspace{5.5cm}
&& \nonumber \\
\hspace{2cm} = 2\qo\,g^2\,[-\varepsilon(\lambda)\cdot \varepsilon(\lambda')]
\frac{\exp(-i\kvec_{2\perp}\cdot\xvec_{2\perp})}{k_{2\perp}^2}
 && \nonumber \\
\widetilde\M[g(q+k_2,\lambda')g(k_1)g(k_3) \to \jpsi(q+k)] \hspace{4cm} &&
\nonumber \\
 = -im\qo\frac{R_0g^5}{m_\perp^4\sqrt{24\pi
m}}
\beta_{T,L}(\lambda') \hspace{2cm}
 && \nonumber \\
 \times \int\frac{d^2\kvec_1}{(2\pi)^2}
\exp(-i\kvec_{1\perp}\cdot \xvec_{1\perp} -i\kvec_{3\perp}\cdot
\xvec_{3\perp}) \frac{\kvec_{1\perp}\cdot \kvec_{3\perp}}{k_{1\perp}^2 \,
k_{3\perp}^2} && \label{mtildedef}
\eeqa

\mysection{Cross sections and $\ptr$-broadening}

In this section we derive the expressions for the production cross sections
and $\ave{\ptr^2}$ of the $\eta_c$ and $\jpsi$ bound states, based on
the amplitudes found in the previous sections.

The differential cross section for scattering on static  centres is
\beq
\frac{d^2\sigma_n}{d^2\Pvec_\perp} = \frac{1}{(2\pi)^2(2\qo)^2} \left\langle
\left\vert {\cal M}_n \right\vert^2 \right\rangle \ ,  \label{fourone}
\eeq
where ${\cal M}_n$ is the bound state production amplitude induced by $n$
scatterings $(n\ge 1)$ and $\Pvec_\perp= 2\pvec_{1\perp}$. In addition to the
implicit color/spin summation and averaging, the average sign in
\eq{fourone} denotes an average over positions of the scattering centres.

In the case of one scattering centre, and assuming a uniform distribution
of this centre with a constant volume density $\rho$, our description is
equivalent to the standard infinite momentum frame description with a
gluon distribution $\propto 1/x$ (see Appendix A).
For $n>1$ centres we
assume in the following that the centres are distributed independently with
the same constant density $\rho$. Thus the average of \eq{fourone} is
defined to be
\beq
\left\langle{\left\vert {\cal M}_n \right\vert}^2 \right\rangle \equiv
\int\prod_{i=1}^{n}(\rho\,d^3\xvec_i)\, \overline{\left\vert {\cal M}_n
\right\vert^2}     \label{fourtwo}
\eeq
where the overline indicates the remaining color/spin sum and average. We
also define the volume, transverse area and length of the target by
\beqa
V&=& \int d^3\xvec_i  \nonumber \\
S&=& \int d^2\xvec_{i\perp}  \nonumber \\
L&=& \int dx_i^z  \label{fourfour}
\eeqa
Our approach can in principle also deal with a case of {\em correlated}
centres, which might be of phenomenological relevance, as discussed in the
final section.

 From \eq{fourone} we obtain the $\ave{P_\perp^2}$ induced by $n$
scatterings as
\beq
\ave{P_\perp^2}_n= \frac{\int d^2\Pvec_\perp\,\,\Pvec_\perp^2\, d^2\sigma_n/
d^2\Pvec_\perp}{\int d^2\Pvec_\perp\, d^2\sigma_n/ d^2\Pvec_\perp}
\label{fourfive}
\eeq

\subsection{$\eta_c$ production}

{\em One scattering}

We first give the cross section for the basic single gluon exchange process.
Squa\-ring the amplitude $\M(gg\to \eta_c)$ of \eq{ggetac} and using
$\varepsilon = (0,1,\pm i,0)/\sqrt{2}$ for the projectile gluon polarization
states we obtain
\beq
\frac{d\sigma_1}{dP_\perp^2}= \frac{4\pi}{9}\rho V \frac{R_o^2\as^3}
{M(P_\perp^2+M^2)^2 P_\perp^2}\ ,  \label{foursix}
\eeq
where $M=2m$ and the factor $\rho V$ arises from the trivial averaging over
$\xvec_1$. Integrating over $P_\perp^2$ yields
\beq
\sigma_1 = \frac{4\pi}{9}\rho V \frac{R_o^2\as^3}{M^5}\log\left(
\frac{M^2}{\lqcd^2}\right) \ .  \label{fourseven}
\eeq

\medskip

{\em Two scatterings}

For the production process induced by two scatterings we use the convolution
formula of \eq{eta2conv}, which as we now shall show implies a similar
convolution at the level of the cross section,
\beq
\frac{d^2\sigma_2}{d^2\Pvec_\perp}= \int d^2\kvec_{1\perp} \left[
\frac{d^2\Pel}{d^2\kvec_{1\perp}}\right]_1\,
\frac{d^2\sigma_1(\Pvec_\perp -\kvec_{1\perp})}{d^2\Pvec_\perp}
\label{foureight}
\eeq
The first factor is the probability density for gluon elastic scattering off
the first static centre. Thus $d^2\sigma_2/d^2\Pvec_\perp$ is given by a
convolution of the probability for having a gluon elastic scattering of
transfer $\kvec_{1\perp}$ {\em first} and the differential cross section for
the basic production process with a single scattering of momentum
transfer $\Pvec_\perp-\kvec_{1\perp}$ occurring {\em thereafter}.

To derive \eq{foureight} we express the elastic gluon and $\eta_c$
production amplitude in the form
\beqa
i\M_{el}[g(q,\lambda)g(k_1)\to g(q+k_1,\lambda')] &=&
f_{ca_1c'}T_{A_1'A_1}^{a_1} \exp(-i\kvec_{1\perp}\cdot \xvec_{1\perp})
\varepsilon(\lambda)\cdot \varepsilon(\lambda') \nonumber \\
&\times& i\widehat\M_{el}[q,k_1]
\label{fournine} \\
i\M[g(q+k_1,\lambda')g(k_2)\to \eta_c(q+k)] &=& T_{A_2'A_2}^{c'}
\exp(-i\kvec_{2\perp}\cdot \xvec_{2\perp})\varepsilon^\mu(\lambda')
 \nonumber \\
&\times& i\widehat\M_\mu[q+k_1,P]   \label{fourten}
\eeqa
where the expressions for the $\widehat\M$ amplitudes are directly obtained
from Eqs. (\ref{ggetac}) and (\ref{ggel}).

Squaring the amplitude (\ref{eta2conv}) leads to
\beqa
& &\frac{d^2\sigma_2}{d^2\Pvec_\perp} = \frac{1}{(2\pi)^2}\int
\frac{d^2\kvec_{1\perp}}{(2\pi)^2} \frac{d^2\kvec_{1\perp}'}{(2\pi)^2}
\exp\left[-i(\kvec_{1\perp}-\kvec_{1\perp}')\cdot (\xvec_{1\perp}
-\xvec_{2\perp})\right] \nonumber \\
&\times& f_{ca_1c'}f_{ca_1'c''}T_{A_1'A_1}^{a_1}T_{A_1A_1'}^{a_1'}
[\varepsilon(\lambda) \cdot \varepsilon(\lambda')]
[\varepsilon(\lambda) \cdot \varepsilon(\lambda'')]^*
\left(\frac{\widehat\M_{el}[q,k_1]}{2\qo}\right)\,
\left(\frac{\widehat\M_{el}[q,k_1']}{2\qo}\right)^*  \nonumber \\
&\times& T_{A_2'A_2}^{c'}T_{A_2A_2'}^{c''}
\varepsilon^{\mu'}(\lambda') \varepsilon^{\mu''}(\lambda'')^*
\left(\frac{\widehat\M_{\mu'}[q+k_1,P]}{2\qo}\right)\,
\left(\frac{\widehat\M_{\mu''}[q+k_1',P]}{2\qo}\right)^*
 \nonumber \\ \label{foureleven}
\eeqa
Summing and averaging over colors, over positions $\xvec_1,\xvec_2$ and over
the intermediate polarizations $\lambda',\lambda''$ gives
\beq
\frac{d^2\sigma_2}{d^2\Pvec_\perp} = \half\int d^2\kvec_{1\perp}
\left[\frac{1}{(2\pi)^2}\frac{\rho L}{2}
\left\vert \frac{\widehat\M_{el}[q,k_1]}{2\qo} \right\vert^2\right]
\left[\frac{1}{(2\pi)^2}\frac{\rho V}{2N}
\left\vert\frac{\varepsilon^\mu\widehat\M_{\mu}[q+k_1,P]}{2\qo}
\right\vert^2\right]
 \nonumber \\  \label{fourtwelve}
\eeq
The second factor in the integrand is readily checked to give the second
factor in \eq{foureight}. The first factor is related to the probability
density for gluon elastic scattering through
\beqa
\frac{d^2\Pel}{d^2\kvec_{1\perp}} &=& \frac{1}{S}
\frac{d^2\sigma_{el}}{d^2\kvec_{1\perp}} =
\frac{1}{(2\pi)^2}\frac{1}{S}
\left\langle\left\vert\frac{\M_{el}}{2\qo}\right\vert^2\right\rangle
\nonumber \\
&=& \frac{1}{(2\pi)^2}\frac{\rho L}{2}
\left\vert\frac{\widehat\M_{el}}{2\qo}\right\vert^2
=\frac{1}{(2\pi)^2}\frac{\rho L}{2} \frac{g^4}{k_{1\perp}^4}
\label{fourthirteen}
\eeqa

The overall factor $\half$ in \eq{fourtwelve} arises from the constraint
$x_1^z<x_2^z$. The bracket notation $[\ ]_1$ in \eq{foureight} is to be
understood as implying that the gluon elastic scattering occurs {\em before}
the hard transfer $\Pvec_\perp-\kvec_{1\perp}$. Absorbing thus the factor
$\half$ in the definition of the probability density, \eq{foureight} follows
from \eq{fourtwelve}.

\medskip

{\em $n$ scatterings}

The generalization of \eq{foureight} to the case of $n$ scatterings may be
written as
\beq
\frac{d^2\sigma_n}{d^2\Pvec_\perp} = \int\prod_{i=1}^{n-1}
\left(d^2\kvec_{i\perp} \left[\frac{d^2\Pel}{d^2\kvec_{i\perp}}\right]_i
\right)\, \frac{d^2\sigma_1(\Pvec_\perp-\sum_{j=1}^{n-1}\kvec_{j\perp})}
{d^2\Pvec_\perp}   \label{fourfourteen}
\eeq

Using Eqs. (\ref{fourfive}), (\ref{foursix}) and (\ref{fourthirteen}) we
obtain the mean transverse momentum squared in $n$ scatterings,
\beq
\ave{P_\perp^2}_n=
\frac
{\dis
{\int\prod_{i=1}^{n-1}
\left(\frac{d^2\kvec_{i\perp}}{k_{i\perp}^4}\right)\,
\int\frac{d^2\kvec_{n\perp}}{k_{n\perp}^2 M_\perp^4}
\left(\sum_{j=1}^{n}\kvec_{j\perp}\right)^2}
}
{\dis
{\int\prod_{i=1}^{n-1}
\left(\frac{d^2\kvec_{i\perp}}{k_{i\perp}^4}\right)\,
\int\frac{d^2\kvec_{n\perp}}{k_{n\perp}^2 M_\perp^4}}
}
\label{fourfifteena}
\eeq
where the correct value $M_\perp^2$ to use is (see the discussion following
\eq{gggetacfin})
\beq
M_\perp^2 = M^2+P_\perp^2 = M^2+(\kvec_{1\perp}+\ldots +\kvec_{n\perp})^2
\label{fourfifteenb}
\eeq
To logarithmic accuracy, $\log(M^2/\lqcd^2)\gg 1$, the lower and upper
limits of the logarithmic integrals in \eq{fourfifteena} can be set to
$\lqcd^2$ and $M^2$, respectively. The result is
\beq
\ave{P_\perp^2}_n= \frac{M^2}{\log\left(\dis{\frac{M^2}{\lqcd^2}}\right)}
+(n-1)\lqcd^2 \log\left(\dis{\frac{M^2}{\lqcd^2}}\right)   \label{foursixteen}
\eeq
We know from section 2 that in $\eta_c$ production induced by $n$
scatterings the last transfer $\kvec_{n\perp}$ is hard, whereas the $(n-1)$
first ones are soft. Thus we may write \eq{foursixteen} in a transparent
way, denoting $n_{soft}=n-1$,
\beq
\ave{P_\perp^2}_n= \sum_{i=1}^n \ave{k_{i\perp}^2} =
\ave{k_{n\perp}^2} + n_{soft} \ave{k_{soft\,\perp}^2}
\label{fourseventeen}
\eeq

It is interesting to relate the mean value of $n_{soft}$ to $\Pel$.
Integrating \eq{fourfourteen} over $\Pvec_\perp$ we get
\beq
\sigma_n= \left[\Pel\right]_1 \cdots \left[\Pel\right]_{n-1} \sigma_1
= \frac{1}{n!} \Pel^{n-1} \sigma_1   \label{foureighteen}
\eeq
The measured $\ave{P_\perp^2}$ is obtained by summing over the number of
rescatterings $n$,
\beqa
\ave{P_\perp^2}&=& \frac
{\sum_{n=1}^\infty \ave{P_\perp^2}_n \sigma_n}
{\sum_{n=1}^\infty \sigma_n} =
\frac
{\sum_{n=1}^\infty \left(\ave{P_{\perp}^2}_1 + n_{soft}
\ave{k_{soft\,\perp}^2}\right) \sigma_n}
{\sum_{n=1}^\infty \sigma_n}  \nonumber \\
&=& \ave{P_{\perp}^2}_{n=1} + \ave{n_{soft}}\ave{k_{soft\,\perp}^2}
\label{fournineteen}
\eeqa
For $\Pel\ll 1$ we have
\beq
\ave{n_{soft}} \equiv \ave{n-1}= \frac
{\sum_{n=2}^\infty (n-1)\sigma_n}
{\sum_{n=1}^\infty \sigma_n} \simeq \frac{\sigma_2}{\sigma_1}
= \frac{\Pel}{2} \ll 1 \ \>.   \label{fourtwenty}
\eeq
Since $\ave{n_{soft}}$ is given by the fraction of events where a
single soft scattering precedes the hard one it is proportional to
the length of the target. This soft scattering is responsible for the
medium-induced $P_{\perp}^2$ of the $\eta_c$,
\beq
\Delta P_\perp^2 = \half \Pel \ave{k_{soft\,\perp}^2}
\label{fourtwentyone}
\eeq
which, according to \eq{foursixteen}, depends on the mass $M$ of the
bound state\footnote{We have assumed a fixed coupling constant $\alpha_s$.
Taking
into  account the running of $\alpha_s$ just replaces the logarithmic
dependence
of the medium-induced $P_{\perp}^2$ in \eq{foursixteen} by a {\it double}
logarithmic dependence.}.
The derivation of the $P_\perp$-broadening of the
$\jpsi$ proceeds along the same lines, which we now summarize.

\subsection{$\jpsi$ production}

As we saw in section 3, when gluon radiation is neglected the basic $\jpsi$
production process involves two hard scatterings. We quote only the
transverse $\jpsi$ cross section obtained from Eqs.
(\ref{gggjpsifin}) and (\ref{fourone}),
\beq
\sigma_2 = \frac{40\pi}{81} \rho^2 VL \frac{R_o^2\as^5}{M^7}
\log^2\left(\frac{M^2}{\lqcd^2}\right)  \label{fourtwentytwo}
\eeq

The cross section for three scatterings is obtained from the amplitude
given by \eq{4gjpsifact}. Since $d_{ca_1d}f_{ca_1e}=0$, the interference
between $\M_1$, where $\kvec_{1\perp}$ is hard, and $\M_2$, where
$\kvec_{2\perp}$ is hard, vanishes and one gets a similar convolution as
\eq{foureight} for the $\eta_c$,
\beqa
\frac{d^2\sigma_3}{d^2\Pvec_\perp}&=& \frac{1}{3}\int d^2\kvec_{2\perp}
\frac{d^2\Pel}{d^2\kvec_{2\perp}}\
\frac{d^2\sigma_2^{13}(\Pvec_\perp -\kvec_{2\perp})}{d^2\Pvec_\perp}
\nonumber \\
&+& \frac{1}{3}\int d^2\kvec_{1\perp}
\frac{d^2\Pel}{d^2\kvec_{1\perp}}\
\frac{d^2\sigma_2^{23}(\Pvec_\perp -\kvec_{1\perp})}{d^2\Pvec_\perp}
\label{fourtwentythree}
\eeqa
The upper indices of $\sigma_2$ indicate which transfers are hard and the
factor $1/3$ is due to the fact that the integration over longitudinal
positions yields $1/2!$ for $\sigma_2$ and $1/3!$ for $\sigma_3$. Absorbing
this factor by defining the probability density to have the (soft) elastic
scattering on a given centre $i$, $1\le i\le 3$, we find
\beqa
\frac{d^2\sigma_3}{d^2\Pvec_\perp}&=& \int d^2\kvec_{2\perp} \left[
\frac{d^2\Pel}{d^2\kvec_{2\perp}}\right]_2\
\frac{d^2\sigma_2^{13}(\Pvec_\perp -\kvec_{2\perp})}{d^2\Pvec_\perp}
\nonumber \\
&+& \int d^2\kvec_{1\perp} \left[
\frac{d^2\Pel}{d^2\kvec_{1\perp}}\right]_1\
\frac{d^2\sigma_2^{23}(\Pvec_\perp -\kvec_{1\perp})}{d^2\Pvec_\perp}
\label{fourtwentyfour}
\eeqa

\eq{fourtwentyfour} generalizes in a straightforward way to $n$ scatterings,
\beq
\frac{d^2\sigma_n}{d^2\Pvec_\perp}= \sum_{i=1}^{n-1} \int
\prod_{\buildrel {j=1} \over {j\neq i}}^{n-1} \left(
d^2\kvec_{j\perp}\left[
\frac{d^2\Pel}{d^2\kvec_{j\perp}}\right]_j \right)\
\frac{d^2\sigma_2^{in}(\Pvec_\perp -\sum_{j=1}^{n-1}\kvec_{j\perp}+
\kvec_{i\perp})}{d^2\Pvec_\perp}  \label{fourtwentyfive}
\eeq
where $i$ and $n$ denote the hard transfers, while the $n-2$ other transfers
labelled by $j$ are soft. \eq{fourtwentyfive} is valid for both transverse
and longitudinal $\jpsi$.

The expression for $\ave{P_\perp^2}_n$ is then
\beq
\ave{P_\perp^2}_n= \frac
{\dis{\int\prod_{i=1}^{n}
\left(\frac{d^2\kvec_{i\perp}}{k_{i\perp}^4}\right)\,
\left[(\kvec_{1\perp}\cdot\kvec_{n\perp})^2+ \ldots
+(\kvec_{{n-1}\perp}\cdot\kvec_{n\perp})^2 \right]
\frac{\beta_{T,L}^2}{M_\perp^8}
\left(\sum_{j=1}^{n}\kvec_{j\perp}\right)^2}}
{\dis{\int\prod_{i=1}^{n}
\left(\frac{d^2\kvec_{i\perp}}{k_{i\perp}^4}\right)\,
\left[(\kvec_{1\perp}\cdot\kvec_{n\perp})^2+ \ldots
+(\kvec_{{n-1}\perp}\cdot\kvec_{n\perp})^2 \right]
\frac{\beta_{T,L}^2}{M_\perp^8}}}  \label{fourtwentysix}
\eeq
where
\beqa
\beta_T^2 &=& 1  \nonumber \\
\beta_L^2 &=& \frac{4M^2 P_\perp^2}{M_\perp^4} =
\frac{4M^2}{M_\perp^4} \left(\sum_{j=1}^{n}\kvec_{j\perp}\right)^2
\label{fourtwentyseven}
\eeqa
In \eq{fourtwentysix} it is necessary (for longitudinally
polarized $\jpsi$'s) to keep the full expression for $M_\perp^2$ which
appears in \eq{4gjpsifin}.

For transversally polarized $\jpsi$'s we find, similarly to the case of the
$\eta_c$,
\beqa
\ave{P_\perp^2}_n &=& 2\ave{k_{hard\,\perp}^2} +
(n-2)\ave{k_{soft\,\perp}^2}  \nonumber \\
&=& \frac{2}{3}\frac{M^2}{\log\left(\dis{\frac{M^2}{\lqcd^2}}\right)}
+(n-2)\,\lqcd^2 \log\left(\frac{M^2}{\lqcd^2}\right)
\label{fourtwentyeight}
\eeqa
Averaging over the number of scatterings $n$, and denoting $n_{soft}=n-2$,
\beq
\ave{P_\perp^2} = \ave{P_\perp^2}_{n=2}+ \ave{n_{soft}} \ave{k_{soft\,\perp}^2}
\label{fourtwentynine}
\eeq
 From \eq{fourtwentyfive} we have
\beq
\sigma_n = 2\frac{n-1}{n!} \Pel^{n-2} \sigma_2  \label{fourthirty}
\eeq
which for $\Pel \ll 1$ yields
\beq
\ave{n_{soft}} \equiv \ave{n-2} = \frac{\sum_{n=3}^\infty (n-2)\sigma_n}
{\sum_{n=2}^\infty \sigma_n} \simeq \frac{\sigma_3}{\sigma_2} =
\frac{2}{3}\Pel \ll 1   \label{fourthirtyone}
\eeq
The medium-induced $\ave{P_\perp^2}$ for transversally polarized $\jpsi$ is
thus
\beq
\Delta P_\perp^2 = \frac{2}{3}\Pel \ave{k_{soft\,\perp}^2}
\label{fourthirtytwo}
\eeq
and depends logarithmically on $M$. We note that $ \frac{2}{3}\Pel$ is
the probability to have one soft gluon scattering before the second (and
last) hard transfer.

In the case of longitudinally polarized $\jpsi$ the expression for
$\ave{P_\perp^2}$ is more involved since the second term in
\beq
\ave{P_\perp^2}_n = \sum_i \ave{\kvec_{i\perp}^2} +
2\sum_{i<j}\ave{\kvec_{i\perp} \cdot \kvec_{j\perp}}
\eeq
is non-vanishing. However, we find that $\Delta P_\perp^2$ still depends
logarithmically on $M$.

\mysection{Discussion}

In the present work we studied rescattering effects in hard collisions.
Most PQCD calculations so far have concentrated on the hard vertex itself,
thus involving only a single projectile and target parton. We were
particularly motivated by the discrepancies observed between theory and data
for quarkonium hadroproduction, and the
evidence for large nuclear effects both
in the cross section and in the average transverse momentum.

To facilitate the calculations we assumed the target scattering to occur off
fixed centres, which selects Coulomb exchange. As illustrated in the
Appendix, at least for a single scattering this is equivalent, in the high
energy limit $(x_{targ} \to 0)$, to a standard calculation with a specific
target gluon structure function. Since the assumption of Coulomb exchange
simplifies the calculations considerably, it will be worthwhile to
investigate how general this equivalence in fact is.

As we already discussed in the Introduction, our calculation shows how soft
scattering factorizes from the hard vertex. This is particularly delicate in
the case of $\jpsi$ production, where soft scattering may occur between the
two hard exchanges, and thus affect the color structure of the hard vertex
itself. An unexpected result of the calculation was that the (relatively)
large longitudinal momentum transfer required to put the heavy quark pair on
its mass shell always occurs at the first scattering, even if that
scattering is soft in terms of the transverse momentum exchanged. This
implies that our analysis is applicable to fixed target data on $\Upsilon$
production, even though the beam energy is not large enough for the $b\bar b$
pair to travel very far off its mass-shell.
Note that these general features remain true when considering
$\jpsi$ or $\Upsilon$ photoproduction in our model.

The exchange of two hard gluons, required for radiationless $\jpsi$ production,
is typically a higher twist effect and thus is
suppressed by a power of the
heavy quark mass $m_Q$. This is because of the small probability to find two
partons within a transverse distance of order $1/m_Q$ from the heavy quarks.
However, our calculation focusses attention on the fact that there actually
are partons within such close distance, namely those created by the $Q^2$
evolution of the projectile and target partons. Thus the projectile gluon of
Fig. \ref{genfig} in reality is not on-shell, but has a logarithmically
distributed virtuality due to previously emitted gluons.
It is `hard'
in the sense of \eq{khard}, and due to its relatively short life-time
must still be at
a short transverse distance from its radiated partners. Could gluon
exchange between the quark pair and such partners
be an important effect in $\jpsi$ formation?

Such a contribution is in fact nothing but a higher order loop effect in the
$\jpsi$ production amplitude, and hence is definitely of leading twist. It
is suppressed by a power of $\as$ in the cross section compared to the
lowest order process involving gluon emission. However, this is at least
partly compensated by the fact that no energy is lost from the quark pair.
Such exchanges could thus be particularly important for the Tevatron data on
$\jpsi$ production at large $\ptr$, due to the strong `trigger bias' effect
which favors hard fragmentation mechanisms. Although a full higher order
calculation of $\jpsi$ hadroproduction is a challenge that remains to be met,
the particular contribution from rescattering off evolutionary gluons should
not be difficult to estimate.

\vspace{1.5 cm}
\noindent
{\bf\large Acknowledgement}

We are grateful for valuable discussions with S. J. Brodsky, M. Cacciari, S.
Gupta, M. Strikman and M. V\"anttinen.

\vskip 20pt
\noindent
\appendix
\noindent
{\bf\LARGE Appendix}\par
\myappendixsection{Relating the heavy target quark density
to the gluon distribution}
In this Appendix we wish to demonstrate the equivalence between our heavy
quark target model and the standard structure function
formulation. In par\-ti\-cu\-lar, we derive the relation between the
number of heavy quarks and the gluon distribution
$G(x)$ in the limit $x \to 0$. We choose to consider the process $\gamma g
\to \qpair$ (see Fig. \ref{figureA1}), where
massless quarks are produced at large $\ptr$. We
expect the equivalence to hold similarly for any other hard process in the
small $x$ limit.

\begin{figure}[htb]
\centerline{\vbox{\epsfxsize=14.5truecm\epsfbox{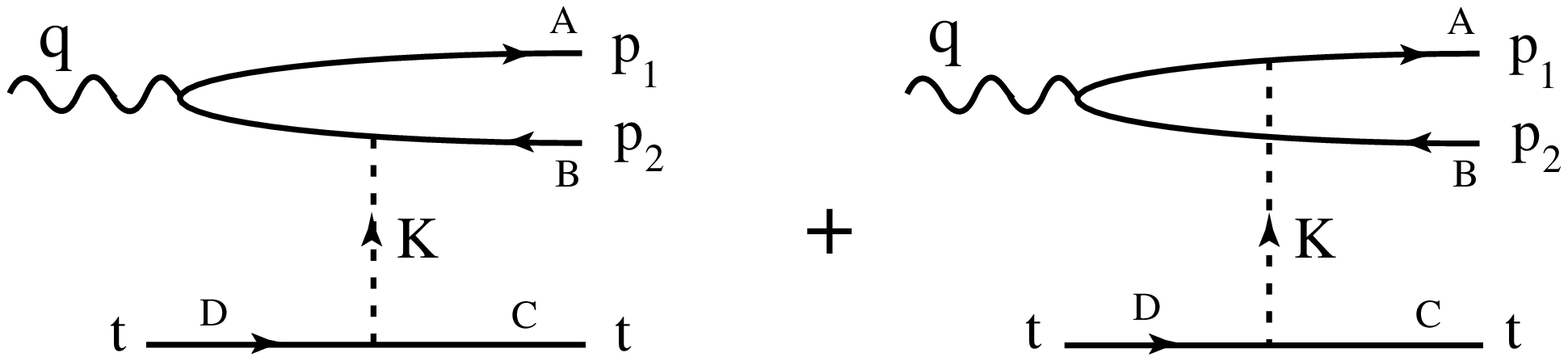}}}
\vspace{2ex}
\caption[*]{Amplitude for the process $\gamma g \to \qpair$.}
\label{figureA1}
\end{figure}

The standard infinite momentum frame expression for the $\gamma g\to \qpair$
cross section in the limit where the photon energy $\nu \to\infty$ while the
transverse momentum $\ptr$ of the produced quarks and their squared invariant
mass $\hat s=xs = 2x\nu m_N$ are kept fixed is
\beq
\frac{d\sigma}{d\ptr^2\,dz}= \frac{\pi e_q^2\alpha\as}{\ptr^4} xG(x)
\left[z^2+(1-z)^2 \right] \label{ggfusion}
\eeq
where $\ptr^2=z(1-z)\hat s$.

In the target rest frame (see Fig. \ref{figureA1})
$z$ is the fraction of the photon
longitudinal momentum carried by the quark,
\beqa
\pvec_1 &=& (\pvec_{1\perp},z\nu) \nonumber \\
\pvec_2 &=& (\pvec_{2\perp},(1-z)\nu+k^z)\\
2\nu k^z &=& -\pvec_{1\perp}^{\ 2}/z - \pvec_{2\perp}^{\ 2}/(1-z) \nonumber
\label{param}
\eeqa
The amplitude for photon scattering on a heavy quark $t$ in the limit we
consider can then be written \cite{BHM}
\beqa
T(\gamma t \to \qpair t) &=& ee_q g^2
T_{AB}^a T_{CD}^a
\frac{4m_t\nu z(1-z)}{\ktr^2} \nonumber \\
&\times& \left[ \frac{\bar u(p_1) \vec \varepsilon \cdot \vec \gamma
v(p_2-k)}{p_{1\perp}^2}-
\frac{\bar u(p_1-k) \vec \varepsilon \cdot \vec \gamma
v(p_2)}{p_{2\perp}^2} \right]
\label{tampl}
\eeqa

After summing and averaging over spins the square of the amplitude
simplifies to
\beq
\overline{|T|^2} = 16 e_q^2\alpha\as^2 (4\pi)^3 \frac{N^2-1}{8N}
\frac{4m_t^2\nu^2 z(1-z)}{\ktr^2 p_{1\perp}^2p_{2\perp}^2}
\left[z^2+(1-z)^2 \right] \label{fusionrest}
\eeq
In the expression for the cross section
\beq
\frac{d\sigma}{dp_{1\perp}^2\,dz}= \frac{1}{4m_t^2(4\pi)^4}
\int \frac{d^2 \vec k_\perp}{z(1-z)\nu^2}\,\overline{|T|^2}  \label{sigexpre}
\eeq
the $\ktr$-dependence of the integrand is (for $\ktr\ll p_{1\perp},\
p_{2\perp}$) of the form $1/\ktr^2$.
Here we introduce $R^{-1} \sim \Lambda_{QCD}$ as an infrared
cut-off for the logarithmic integral, where $R$ is the nucleon
radius, playing the role of a color screening length.
To leading logarithmic
accuracy the scattering occurs incoherently over the pointlike target quarks.
The upper limit of the $\ktr$-integration is
set by $\ktr \simeq p_{1\perp}$, where the approximation
$\vec p_{2\perp}^{\ 2} = (\vec k_\perp - \vec p_{1\perp})^{2} \simeq \vec
p_{1\perp}^{\ 2}$ fails and the integrand is more strongly damped in $\ktr$.
This defines the logarithmic approximation as
\beq
\log\left(\frac{\ptr^2}{\lqcd^2} \right) \gg 1
\label{logappr}
\eeq

In order to deal with measurable cross sections
one has to average over the possible positions
of the heavy target quark.
We assume in this paper a uniform density for the target
quark, denoted by $\rho$, and thus we multiply \eq{sigexpre}
by a factor $\rho V$ counting the number of static centres
in one hard scattering, $V$ being the target volume.

Comparing the resulting
expression for the $\ktr$-integrated cross section
with that of the infinite momentum frame result
(\ref{ggfusion}) we find
\beq
xG(x) =
\rho V\,\frac{N^2-1}{2\pi N} \as(\ptr^2) \log\left(
\frac{\ptr^2}{\lqcd^2} \right)
\label{xG}\ \>,
\eeq
or, taking $N=3$,
\beq
\rho V = \frac{33-2n_f}{16} xG(x)
\label{rhoV}
\eeq

We have checked that the
relation (\ref{rhoV}) is obtained also if one
considers $\eta_c$ production through
two-gluon fusion, by simply comparing the expressions
for the cross section in the target rest frame
obtained from \eq{ggetac} with that of
the infinite momentum frame (see, \eg, Ref. \cite{BaierRueckl}).

\end{document}